\documentclass[12pt]{article}
\usepackage{graphicx}

\newcommand\pubnumber{SNSN-323-63}
\newcommand\pubdate{\today}

\def\napoli{Department of Physics and Astronomy\\
Rice University, Houston, TX, USA}
\def\support{\footnote{Work supported by the Office of Science, 
          Department of Energy, under grant DE-F902-10ER41661 }}

\def\Title#1{\begin{center} {\Large #1 } \end{center}}
\def\Author#1{\begin{center}{ \sc #1} \end{center}}
\def\Address#1{\begin{center}{ \it #1} \end{center}}

\newcommand\pubblock{\rightline{\begin{tabular}{l} \pubnumber\\
         \pubdate  \end{tabular}}}
\newenvironment{Abstract}{\begin{quotation}  }{\end{quotation}}
\newenvironment{Presented}{\begin{quotation} \begin{center} 
             PRESENTED AT\end{center}\bigskip 
      \begin{center}\begin{large}}{\end{large}\end{center} \end{quotation}}
\def\Acknowledgements{\bigskip  \bigskip \begin{center} \begin{large}
             \bf ACKNOWLEDGEMENTS \end{large}\end{center}}

\def\beq{\begin{equation}}
\def\eeq#1{\label{#1}\end{equation}}
\def\eeqn{\end{equation}}

\def\beqa{\begin{eqnarray}}
\def\eeqa#1{\label{#1}\end{eqnarray}}
\def\eeqan{\end{eqnarray}}

\let\bar=\overbar

\def\Dslash{\not{\hbox{\kern-4pt $D$}}}
\def\dslash{\not{\hbox{\kern-2pt $\del$}}}

\def\msb{{\bar{\ssstyle M \kern -1pt S}}}

\begin{document}
\begin{titlepage}
\pubblock

\vfill
\Title{The RHIC Beam Energy Scan Phase II: Physics and Upgrades}
\vfill
\Author{ David Tlusty\support}
\Address{\napoli}
\vfill
\begin{Abstract}
The exploration of the QCD phase diagram has been one of the main drivers of contemporary nuclear physics. The Relativistic Heavy Ion Collider (RHIC) at BNL is uniquely suited for this task through its Beam Energy Scan (BES) program which allowed for a large range in baryon chemical potential $\mu_B$ as was successfully demonstrated after the completion of Phase 1 in 2014. Phase 2 of the BES at RHIC is scheduled to start in 2019 and will explore with precision measurements the intermediate-to-high $\mu_B$ region of the QCD phase diagram, five energies $\sqrt{s_{NN}}$ from 7.7 to 19.6 GeV in  collider mode and eight energies $\sqrt{s{_{NN}}}$ from 3.0 to 7.7 GeV in fixed-target mode. Some of the key measurements are: the net-protons kurtosis that could pinpoint the position of a critical point, the directed flow that might prove a softening of the EOS, and the chiral restoration in the dielectron channel. These measurements will be possible with an order of magnitude better statistics provided by the electron cooling upgrade of RHIC and with the detector upgrades planned to improve STAR's acceptance. These proceedings review the BES Phase-2 program and the physics opportunities enabled by these upgrades.
\end{Abstract}
\vfill
\begin{Presented}
CIPANP 2018\\
Palm Springs, CA, USA,  May 29 - June 3, 2018
\end{Presented}
\vfill
\end{titlepage}
\def\thefootnote{\fnsymbol{footnote}}
\setcounter{footnote}{0}

\section{Introduction}

\begin{figure}[htb]
\centering
\includegraphics[width=0.7\textwidth]{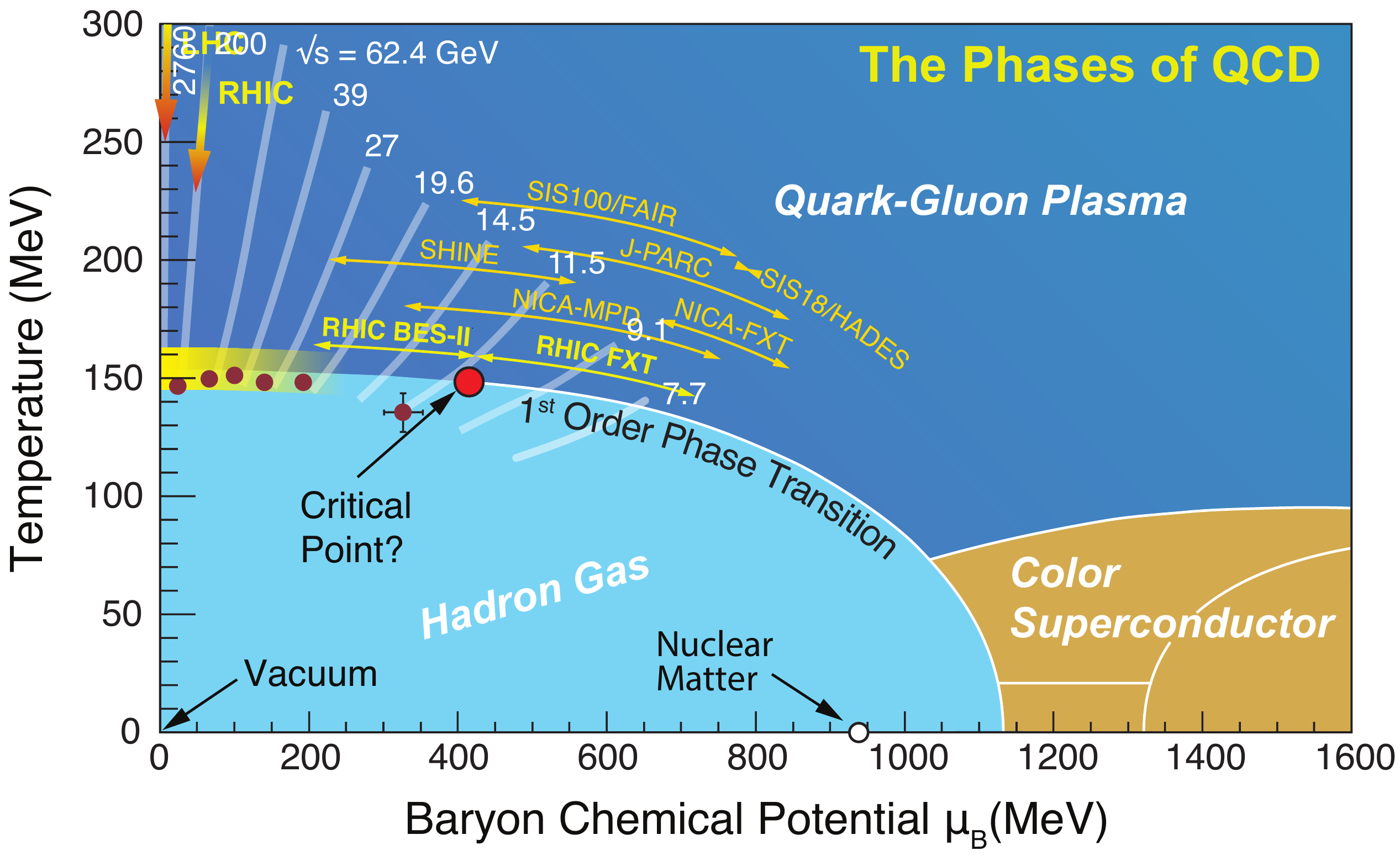}
\caption{A conceptual QCD phase diagram, showing the relevant regions in the plane of temperature versus $\mu_B$. 
Various details, especially the position of the critical point, still remain uncertain. Figure taken from Ref. \cite{HelenQM}}
\label{fig:PhaseDiagram}
\end{figure}

The first decade of operation of the Relativistic Heavy Ion Collider (RHIC) 
was dedicated mostly to Au+Au collisions
at $\sqrt{s_{NN}}=200$ GeV.
Those measurements accumulated enough evidence in support of a deconfined partonic phase
on nuclear matter in the early stages of heavy ion collisions when 
temperatures are higher than the critical temperature $T_c$ 
\cite{ExpSummary1,ExpSummary2,ExpSummary3,ExpSummary4}.
The deconfined matter, when interactions among quark and gluons 
are weak enough due to asymptotic freedom, is called the Quark Gluon Plasma (QGP). 
Finite temperature lattice Quantum Chromodynamics (QCD) calculations predict 
\cite{Aoki} a cross-over from a hadronic to a QGP phase at a vanishing 
baryon chemical potential $\mu_B$ and $T_c=154\pm 9$ MeV
\cite{QCDCritTemp}. 
Several QCD-based calculations \cite{Ejiri,Kapusta} show that at lower $T$ 
and higher $\mu_B$ a first-order phase transition may take place. 
The point in the QCD phase diagram, where the first-order phase transition ends, 
is the QCD critical point \cite{Gupta}. Our current understanding of  
nuclear matter is illustrated by a conceptual phase diagram shown in Fig. \ref{fig:PhaseDiagram}. 
It depicts the estimated position of the critical point at $\mu_B=400$ MeV, but 
we cannot exclude the possibility that a crossover extends to even higher 
baryon chemical potentials. 

The second decade of RHIC operations included the first phase of the 
Beam Energy Scan (BES-I) program with the intent to map out the QCD phase diagram, whose goals are \cite{BES2}
\begin{enumerate}
\item to investigate the expected turn-off of QGP signatures
that have been established at $\sqrt{s_{NN}}=200$ GeV. 
\item to search for the predicted first-order phase transition \cite{Ejiri, Kapusta}
between the hadronic and QGP phase. A promising observable for this is 
the directed flow since it is a proxy for the pressure in a hydrodynamic picture.
Some models \cite{Rischke, Stocker} then predict a dramatic drop in the pressure
equivalent to a softening of the Equation of State (EoS) when a first-order phase
transition occur. 
\item to search for a critical end point, which is expected to exist if the 
QCD phase diagram includes a first-order phase transition. A macroscopic system
at the critical point would show observable effects such as critical opalescence
and hence the characteristic length scale (the correlation length) would become
infinite. Theory suggests \cite{Stephanov} that even in small systems such
as in heavy-ion collisions, the local increase in fluctuations near the 
critical point could be observed experimentally.  
\item to search for chiral symmetry restoration which affects mass and width of the $\rho(770), \omega(782),$ and $\phi(1020)$ vector mesons that can be studied via dilepton decays. 
\end{enumerate}

The end of the second decade of RHIC operations will be dedicated
to a second phase of the BES program (BES-II) program \cite{BES2}. 
Its main goal is to revisit the promising signals from BES-I with greatly 
improved statistics and performance of detectors. Using the RHIC beams
in a fixed-target mode will extend the $\mu_B$ further to 720 MeV. 
The improved physics capability will come partly from three detector upgrades, 
and partly from RHIC beam luminosity increase.    

The STAR detector \cite{STARdetector} has been designed to study polarized 
proton heavy-ion collisions at RHIC. STAR offers uniform acceptance
in azimuthal angle, larger acceptance in transverse momentum in the
mid-rapidity ($-1<y<1$) region, and operability in both collider
and fixed-target mode. Its large mid-rapidity coverage is very well 
suited for the BES physics program.
In the collider mode, changes in STAR's acceptance
could be considered negligible which is crucial especially in 
the fluctuations measurements. STAR is further capable of excellent particle
identification through $dE/dx$ in its Time Projection Chamber (TPC) \cite{TPC}
and Time-of-Flight (TOF) \cite{TOF}. 

These proceedings will discuss some of the remarkable results regarding the
QCD phase diagram exploration from BES-I followed by detector upgrades and
expected improvements in anticipation of BES-II.

\section{Selection of BES-I Measurements}

\subsection{Cumulant Ratios of Net-proton(kaon) Multiplicity}

\begin{figure}[htb]
\centering
\includegraphics[width=0.33\textwidth,clip]{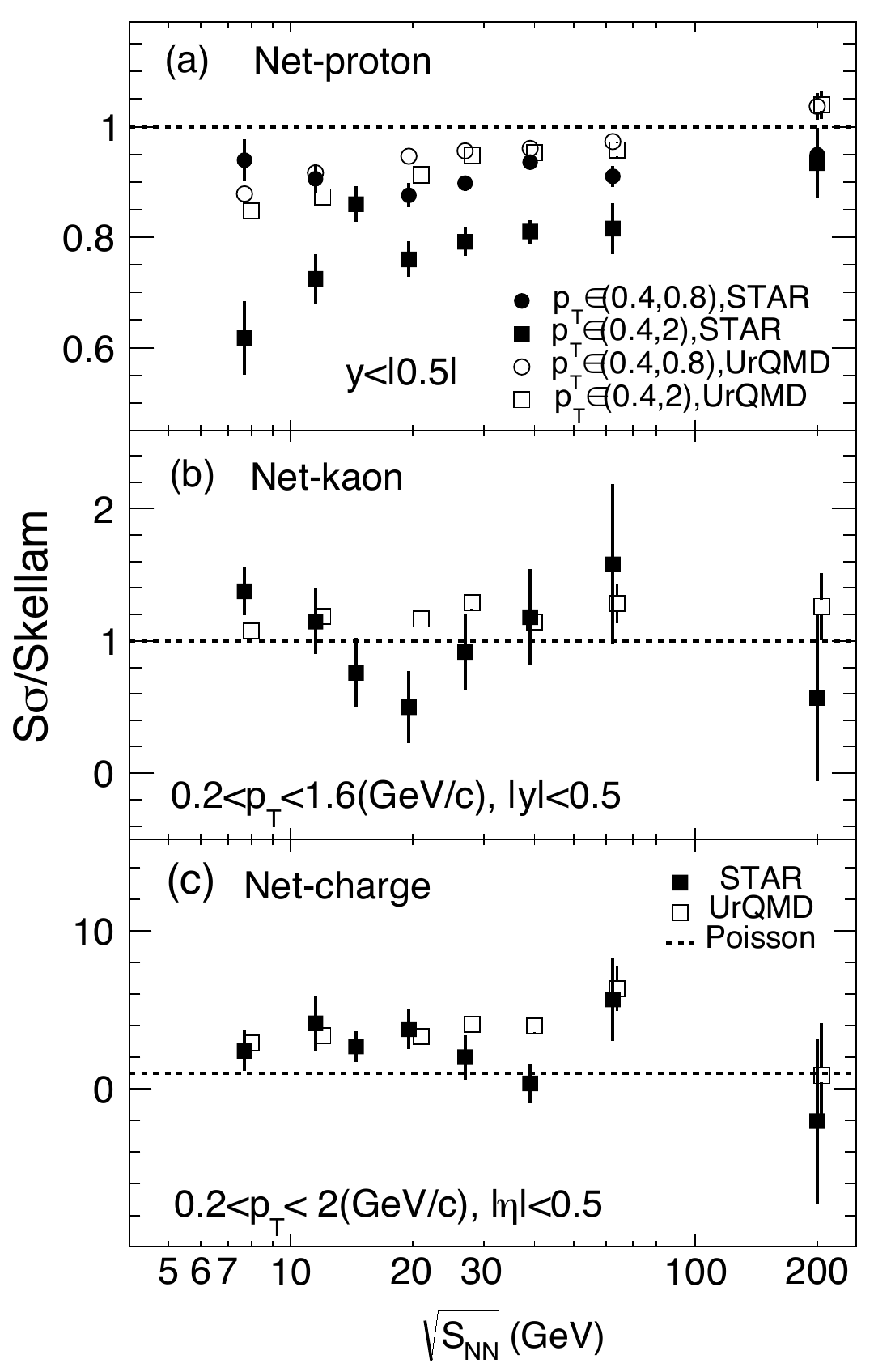}
\includegraphics[width=0.33\textwidth,clip]{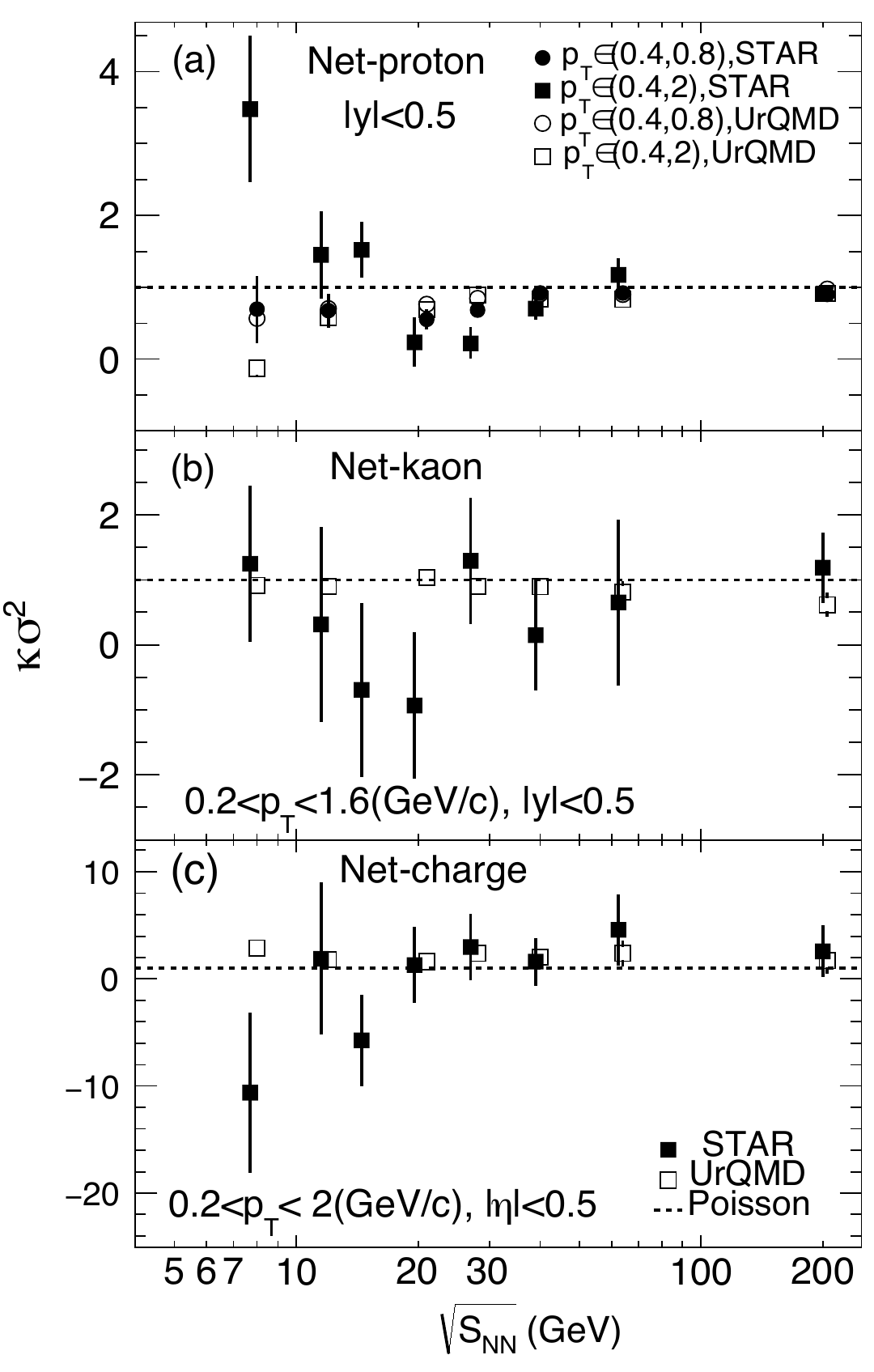}
\caption{Energy dependence of cumulant ratios $S\!\sigma$ (left panel), $\kappa\sigma^2$ (right panel) of net-proton, net-charge, and net-kaon multiplicity distributions for the Au+Au collision at $\sqrt{s_{NN}}=7.7$ to 200 GeV. The solid markers represent the results from the STAR measurement, the open markers represent results from the UrQMD calculation. The dashed lines denote the Poisson expectations for the STAR data. Figure taken from Ref. \cite{Luo}.}
\label{fig-3}
\end{figure}

Fluctuations of conserved quantities, such as baryon, electric charge, and strangeness number, are sensitive observables in the study of phase transitions in the QCD matter and critical point \cite{Luo}. Those fluctuations are studied 
in limited rapidity window which makes them observable \cite{STARacceptFluct}. Statistical moments of multiplicity distributions like skewness,
$S\propto \left\langle (N-\langle N \rangle)^3 \right\rangle$ and kurtosis,
$\kappa \propto \left\langle (N-\langle N \rangle)^4 \right\rangle$ are of particular interest since they are sensitive enough to the correlation length ($\kappa$ to the seventh power for example) given the limited statistics. Furthermore,   
products $\kappa\sigma^2$ and $S\sigma$ are directly related to the ratios of  susceptibilities as
$$ \kappa\sigma^2 = \frac{\chi_4}{\chi_2}\qquad S\sigma = \frac{\chi_3}{\chi_2}$$
which make them readily calculable by lattice QCD. Since the isospin susceptibility remains finite at the critical point, critical fluctuations in net-baryon number are reflected in the fluctuations of the net-proton number \cite{NetBaryonProxy}. Net-kaon number is used as a proxy for net-strangeness
in terms of searching for non-monotonic energy dependence of the fluctuation observable near QCD critical point \cite{NetStrangenessProxy}. 


Figure \ref{fig-3} shows the energy dependence of cumulant ratios $S\!\sigma, \kappa\sigma^2$ of net-proton, net-charge, and net-kaon multiplicity distributions of the 5\% most central Au+Au collisions in RHIC BES-I energies from the STAR experiment \cite{STARcumulants1,STARcumulants2} and UrQMD calculations \cite{Luo}. The non-monotonic energy dependence of the net-proton $\kappa\sigma^2$ and net-kaon $S\!\sigma$ contrasts with the monotonic behavior of 
UrQMD predictions which have not the critical physics implemented. In $\kappa\sigma^2$ of net protons, there is a minimum
below the unity around $\sqrt{s_{NN}}=20$ GeV and a steep rise with further decreasing $\sqrt{s_{NN}}$, ending $\sim$2 standard deviations above unity. 
If $\kappa\sigma^2$ returns to the unity at $\sqrt{s_{NN}}$ below 7.7 GeV, this could suggest a critical point signature of the kind predicted by Stephanov \cite{Stephanov}. 
This result makes a strong case for more data which STAR will collect in BES-II as well as exploration at $\sqrt{s_{NN}}$ below 7.7 GeV which 
will be achieved in the fixed-target mode. 

\subsection{Directed Flow of Net Protons}

Directed flow is defined as the first harmonic coefficient in the Fourier expansion of the azimuthal angle distribution $\phi$ of final-state particles relative to the reaction plane azimuth $\Psi_R$, i.e. $v_1 = \langle \cos(\phi-\Psi_R)\rangle$ \cite{Flow1, Flow2, Flow3}. It can be interpreted as a collective sideward motion of the participant
nucleons in heavy-ion collisions. It is built at early stages of a collision which makes it a very good proxy of 
the pressure in a colliding system, as suggested by nuclear transport \cite{Bass, Bleicher} and hydrodynamic \cite{Hydro}
models. The excitation function of the directed flow slope with respect to rapidity $dv_1/dy$ is predicted to exhibit a minimum at a certain collision energy in hydrodynamical calculations using an EoS with a first-order QCD phase transition  
\cite{Nara}. 

\begin{figure}[htb]
\centering
\includegraphics[width=0.9\textwidth]{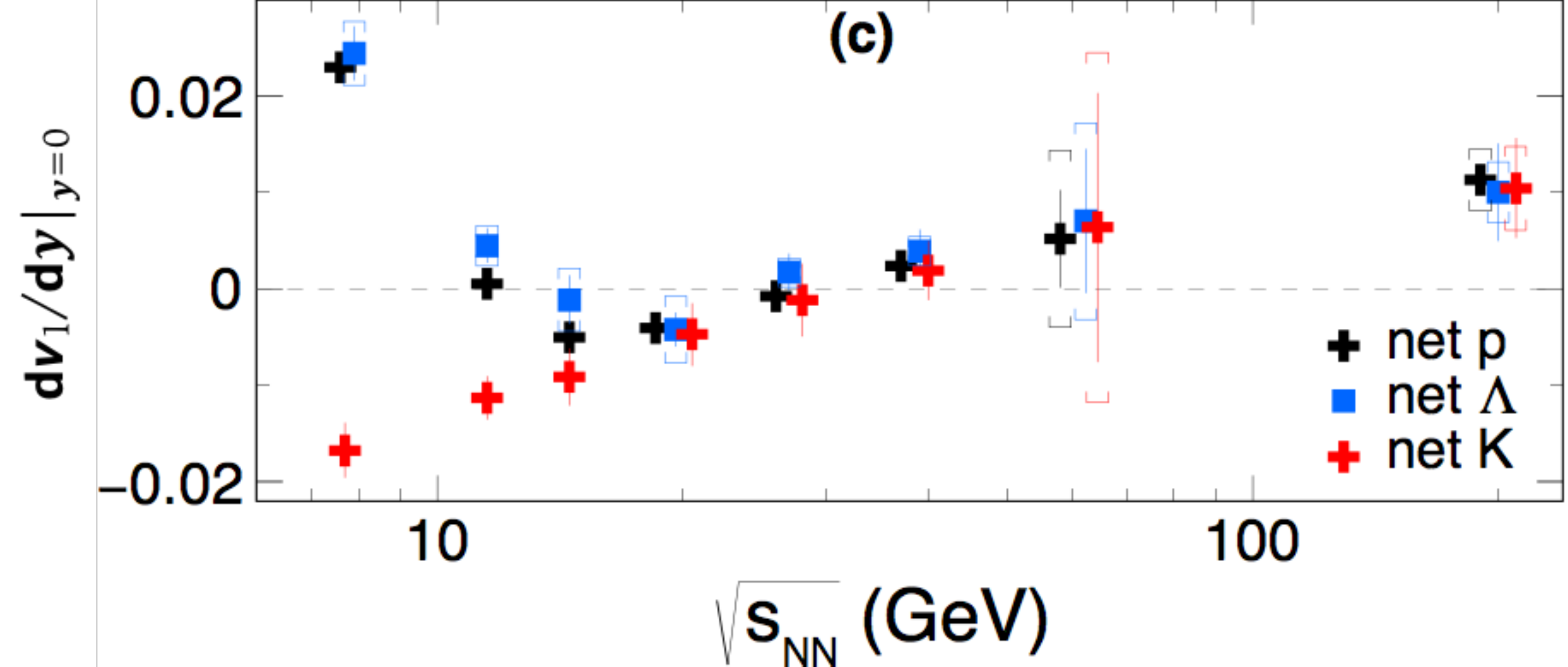}
\caption{Directed flow slope $dv_1/dy$ near midrapidity as a function of $\sqrt{s_{NN}}$ for net protons, net $\Lambda$, and net kaons in 10-40\% centrality Au+Au collisions from STAR. Figure taken from Ref. \cite{StarV1}. }
\label{fig:V1}
\end{figure}

STAR's BES-I measurement of the $dv_1/dy$ of net protons \cite{StarV1}, shown in Fig. \ref{fig:V1}, reveals a non-monothonic behavior of $dv_1/dy$ near mid-rapidity with a minimum between 10 and 20 GeV and double-sign change. Qualitatively, this minimum was predicted by three-fluid hydrodynamic calculations \cite{Rischke, Stocker} only when the EoS has a first-order 
phase transition. However, other theoretical predictions such as Ref. \cite{IvanovSoldatov} are ambiguous on the
issue of whether a crossover or first-order phase transition reproduces STAR's measurements better. The hadronic transport
model JAM \cite{Nara} reports a minimum in the case of a first-order phase transition, but the minimum is about an
order of magnitude more pronouced and lies about a factor 3 lower in $\sqrt{s_{NN}}$. Hence the models need further development 
and experimental constrains from new directed flow measurements in finer centrality bins possible in BES-II.   

\section{Plans and Upgrades for BES-II}
 
\begin{figure}[htb]
\centering
\includegraphics[width=0.46\textwidth,clip]{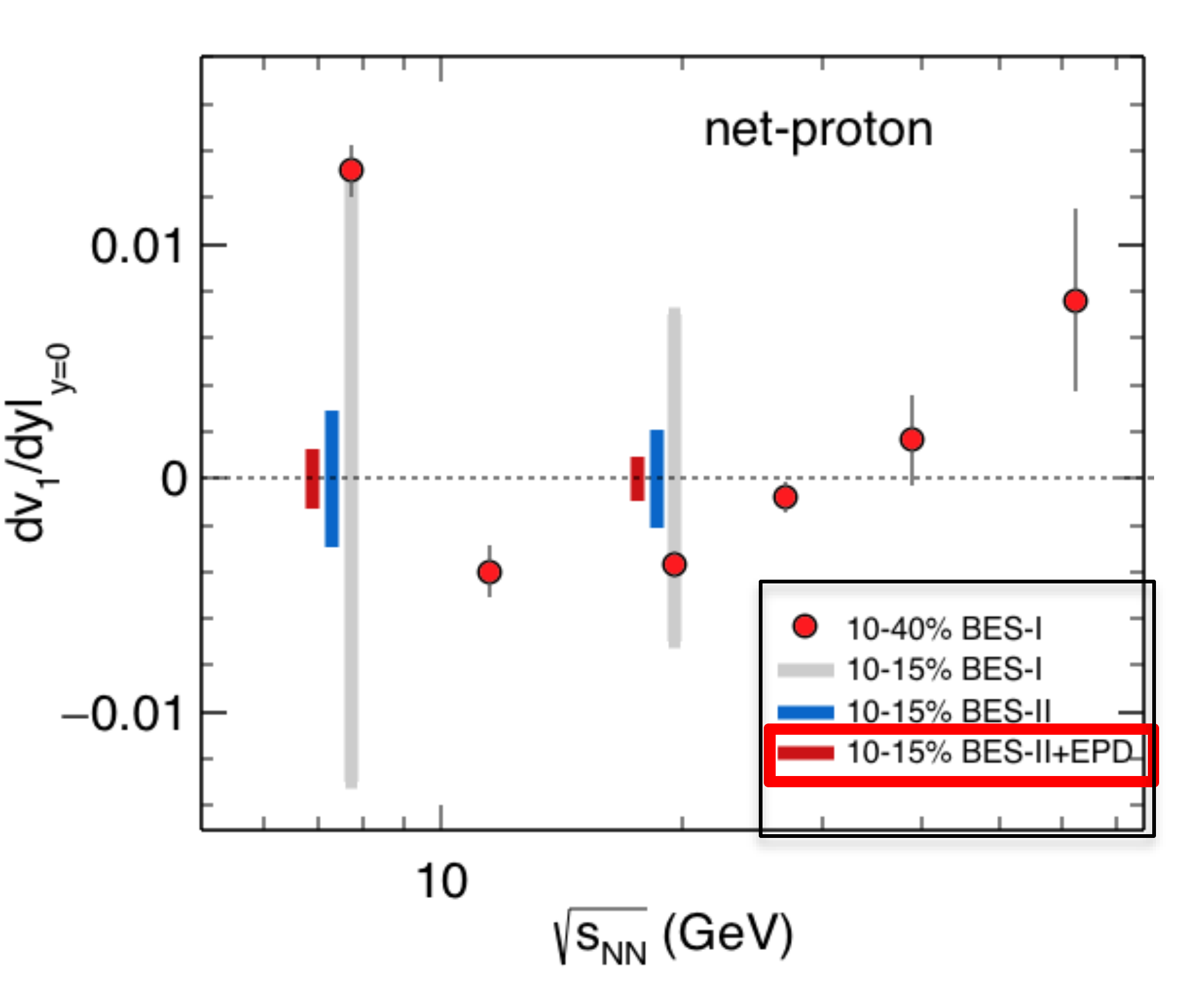}
\includegraphics[width=0.53\textwidth,clip]{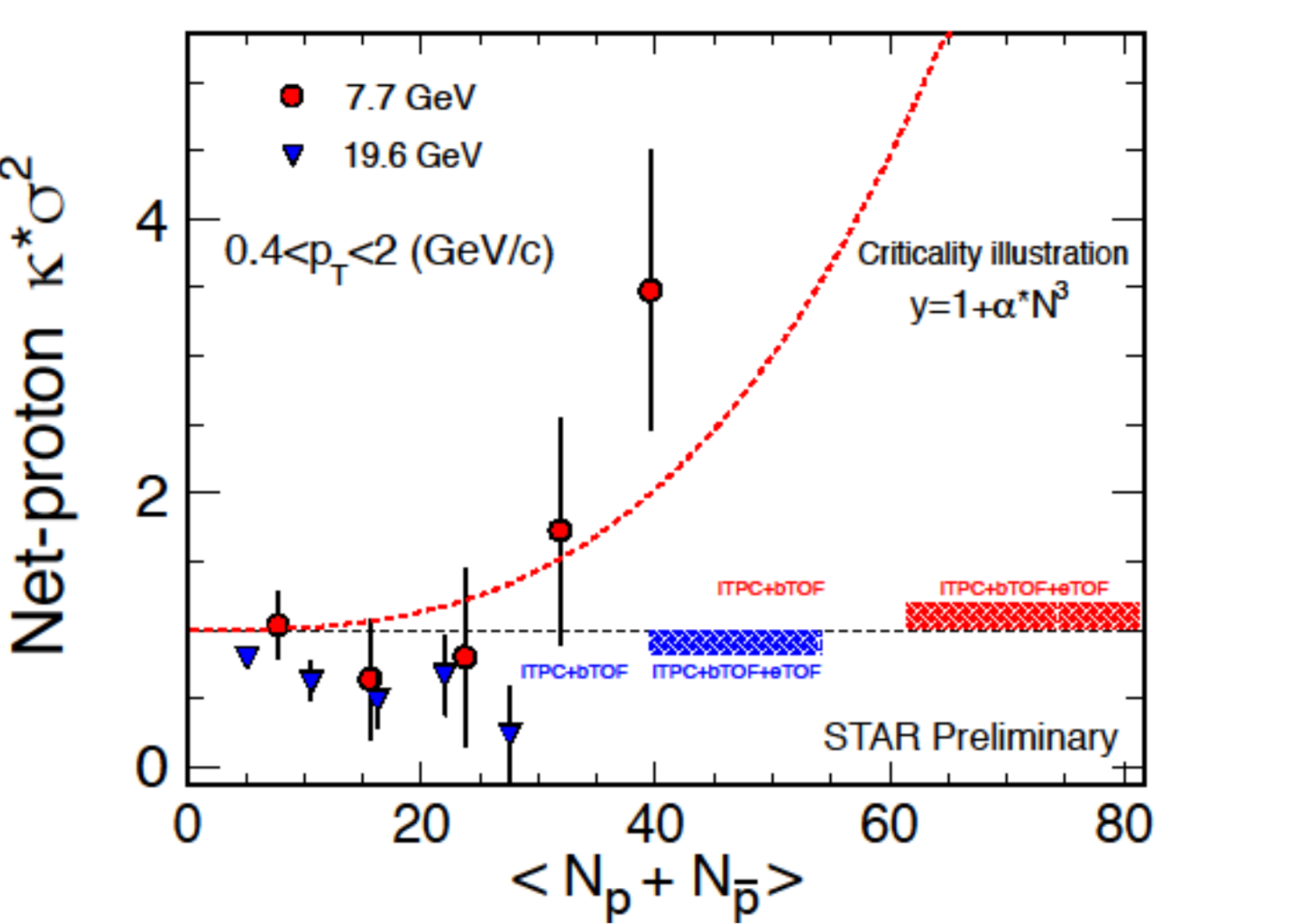}
\caption{Left plot: net-proton directed flow measured by STAR in BES-I \cite{StarV1} (red circles) and systematic error
estimation done at two centre-of-mass energies (7.7 and 19.6 GeV) for BES-I, BES-II and BES-II with EPD upgrade. Plot taken from Ref \cite{EPD}. Right plot: The net-proton $\kappa\sigma^2$ as a function of the proton and anti-proton multiplicity.  Plot taken from Ref \cite{eTOF}.}
\label{BES2improvement}
\end{figure}

Three major STAR detector upgrades have been prepared in anticipation of RHIC BES-II. They are aimed at the  reduction of 
systematic uncertainties and increase of STAR acceptance. At the same time, RHIC will improve its luminosity for low energy
beams to increase statistics at low $\sqrt{s_{NN}}$ and reduce statistical uncertainties.
The following paragraphs list and discuss those upgrades. Figure \ref{BES2improvement} shows the impact
of those upgrades on $dv_1/dy$ (left plot) and net-proton $\kappa\sigma^2$ (right plot). 

The Inner TPC (iTPC) upgrade \cite{iTPC} improves the spatial resolution of the TPC anodes which leads to better particle identification from $dE/dx$ ($\sim$25\% improvement), and better momentum resolution ($\sim$ 15\% improvement at larger momenta). Furthermore, it extends the preudorapidity $\eta$ coverage from $(-1,1)$ to $(-1.5,1.5)$, which also  extends acceptance at low transverse momenta, from $\sim 125$ MeV/$c$ to as low as 60 MeV/$c$.  In addition, the tracking at higher $\eta$ is crucial for the endcap TOF (eTOF) (discussed below).

The Event Plane Detector (EPD) \cite{EPD} is an entirely new subdetector that will improve the event plane
resolution by about a factor of 2 in Au+Au collision at $\sqrt{s_{NN}}=19.6$ GeV and about a factor of 3 at $\sqrt{s_{NN}}=7.7$ GeV. It consists of two azimuthally symmetric scintillator telescopes with high radial and azimuthal segmentation and with a pseudorapidity coverage $2.1<|\eta|<5$ . The EPD will allow the centrality and the event plane to be measured in the forward region, reducing systematics due to autocorrelations from mid-rapidity analyses. The left plot in Fig. \ref{BES2improvement} shows the improvement of net-proton directed flow measurements in BES-II with, and without EPD.

The eTOF \cite{eTOF} is a joint project of STAR and CBM collaborations for the BES-II program. 
It will cover the pseudorapidity region of
$1.1<\eta<1.6$ and will improve particle identification in the $\eta$ acceptance region added by the iTPC. This
will particularly benefit data from fixed-target collisions. 

Net-proton $\kappa\sigma^2$ strongly depends on $p_T$ and rapidity cuts of protons \cite{eTOF}. A new 
approach has been proposed \cite{eTOF}, where $\kappa\sigma^2$ is analysed as a function of the sum of the number of measured protons and anti-protons, as is shown in Fig. \ref{BES2improvement}. The STAR BES-I $\kappa\sigma^2$ for 7.7 GeV trend upward with total protons while for 19.6 GeV, the trend is downward. It is expected that the $\kappa\sigma^2$ signal will be large for energies that create systems near the critical point, while for systems with a baryon chemical potential below the critical point, the $\kappa\sigma^2$ will drop below unity. The added coverage of the eTOF will extend the measurement further in the sum of protons and anti-protons so the fluctuation signal will provide a clearer and more significant indication of critical behavior.

At $\sqrt{s_{NN}}=11.5$ GeV and below, RHIC will use its newly developed Low-energy RHIC Electron Cooler (LEReC)
\cite{Pinayev} which will increase the luminosity about a factor of 4. At $\sqrt{s_{NN}}=14.5$ GeV and above,
accelerator improvements involving bunch structure and $\beta^*$ which will increase the luminosity about a factor of 3.

RHIC luminosity in normal collider mode decreases like relativistic $\gamma^3$ at energies below AGS energies ($\sqrt{s_{NN}}=27$ GeV), because the RHIC ring is used as a decelerator, thus $\sqrt{s_{NN}}=7.7$ was set to be minimum in collider mode. Hence one of the RHIC beams was set to impact a gold fixed target inside the beam pipe at the z-position of one end of the TPC. In 2015, STAR
successfully collected more than 1M good Au+Au events during a 30 min fixed-target test run and demostrated that with little beam time it can reproduce measurements from previous fixed-target experiment, such as E895, E877, and E892 \cite{Kathryn}.

During BES-I running, STAR has collected dielectron data for minimum-bias Au+Au collisions at $\sqrt{s}$ = 19.6, 27, 39, and 62.4 GeV. Below 19.6 GeV, the event size of the data sets have been too small to allow for meaningful 
measurements in the low-mass range of the dielectron spectrum. Estimated 10x higher statistics in BES-II (at $\sqrt{s}$ = 19.6) will allow first measurements at energies below 19.6 \cite{BES2}.


\section{Summary}

We present two of the BES-I goals, namely, the search for a first-order phase transition
through directed flow measurement and critical point through high-moments of net particle multiplicities.
Directed flow of net-protons shows non-monotonic behavior and double sign change. It will be interesting
to study the directed flow of other hadrons ($\Lambda,\overline{\Lambda},K^0_S,K^\pm,\phi$) to disentangle
the role of produced and transported quarks in heavy-ion collisions.
High-moment fluctuations hint at a critical behavior, but need better statistics and extended $\sqrt{s_{NN}}$ 
coverage on the low side, possibly different analysis approach which was referred to in these proceedings. 

BES-II provides a well defined plan for a focused study of the type of a phase transition and localization of the critical point.
Presented detector upgrades will reduce systematic uncertainties and extend kinematical and PID range.
RHIC facility upgrades will increase luminosity and fixed-target program will extend $\mu_B$ and $\sqrt{s_{NN}}$.

\Acknowledgements
This work is part of and supported by U.S. Department of Energy under contract number DE-FG02-10ER41666.

\end{document}